\documentstyle[psfig]{mn}

\newif\ifAMStwofonts

\def\gapprox{{_>\atop{^\sim}}}
\def\lapprox{{_<\atop{^\sim}}}

\def\simlt{\lower.5ex\hbox{$\; \buildrel < \over \sim \;$}}
\def\simgt{\lower.5ex\hbox{$\; \buildrel > \over \sim \;$}}

\ifoldfss
  \ifCUPmtlplainloaded \else
    \NewTextAlphabet{textbfit} {cmbxti10} {}
    \NewTextAlphabet{textbfss} {cmssbx10} {}
    \NewMathAlphabet{mathbfit} {cmbxti10} {} 
    \NewMathAlphabet{mathbfss} {cmssbx10} {} 
  \fi
  \ifAMStwofonts
    \ifCUPmtlplainloaded \else
      \NewSymbolFont{upmath} {eurm10}
      \NewSymbolFont{AMSa} {msam10}
      \NewMathSymbol{\upi}     {0}{upmath}{19}
      \NewMathSymbol{\umu}     {0}{upmath}{16}
      \NewMathSymbol{\upartial}{0}{upmath}{40}
      \NewMathSymbol{\leqslant}{3}{AMSa}{36}
      \NewMathSymbol{\geqslant}{3}{AMSa}{3E}

    \fi
  \fi
\fi 

\ifnfssone
  \newmathalphabet{\mathit}
  \addtoversion{normal}{\mathit}{cmr}{m}{it}
  \addtoversion{bold}{\mathit}{cmr}{bx}{it}
  \newmathalphabet{\mathbfit} 
  \addtoversion{normal}{\mathbfit}{cmr}{bx}{it}
  \addtoversion{bold}{\mathbfit}{cmr}{bx}{it}
  \newmathalphabet{\mathbfss} 
  \addtoversion{normal}{\mathbfss}{cmss}{bx}{n}
  \addtoversion{bold}{\mathbfss}{cmss}{bx}{n}
  \ifAMStwofonts
    \ifCUPmtlplainloaded \else
      \UseAMStwoboldmath
      \makeatletter
      \new@mathgroup\upmath@group
      \define@mathgroup\mv@normal\upmath@group{eur}{m}{n}
      \define@mathgroup\mv@bold\upmath@group{eur}{b}{n}
      \edef\UPM{\hexnumber\upmath@group}
      \new@mathgroup\amsa@group
      \define@mathgroup\mv@normal\amsa@group{msa}{m}{n}
      \define@mathgroup\mv@bold\amsa@group{msa}{m}{n}
      \edef\AMSa{\hexnumber\amsa@group}
      \makeatother
      \mathchardef\upi="0\UPM19
      \mathchardef\umu="0\UPM16
      \mathchardef\upartial="0\UPM40
      \mathchardef\leqslant="3\AMSa36
      \mathchardef\geqslant="3\AMSa3E
    \fi
  \fi
\fi 

\ifnfsstwo
  \DeclareMathAlphabet{\mathbfit}{OT1}{cmr}{bx}{it}
  \SetMathAlphabet\mathbfit{bold}{OT1}{cmr}{bx}{it}
  \DeclareMathAlphabet{\mathbfss}{OT1}{cmss}{bx}{n}
  \SetMathAlphabet\mathbfss{bold}{OT1}{cmss}{bx}{n}
  \ifAMStwofonts
    \ifCUPmtlplainloaded \else
      \DeclareSymbolFont{UPM}{U}{eur}{m}{n}
      \SetSymbolFont{UPM}{bold}{U}{eur}{b}{n}
      \DeclareSymbolFont{AMSa}{U}{msa}{m}{n}
      \DeclareMathSymbol{\upi}{0}{UPM}{"19}
      \DeclareMathSymbol{\umu}{0}{UPM}{"16}
      \DeclareMathSymbol{\upartial}{0}{UPM}{"40}
      \DeclareMathSymbol{\leqslant}{3}{AMSa}{"36}
      \DeclareMathSymbol{\geqslant}{3}{AMSa}{"3E}
    \fi
  \fi
\fi 

\ifCUPmtlplainloaded \else
  \ifAMStwofonts \else 
    \def\upi{\pi}
    \def\umu{\mu}
    \def\upartial{\partial}
  \fi
\fi

\title[Solar Seismic Model as a New Constraint on Supersymmetric Dark Matter]
    {Solar Seismic Model as a New \\ Constraint on Supersymmetric Dark Matter}
\author[Lopes, Bertone \& Silk]
{Il\'\i dio P. Lopes$^{1,2}$, Gianfranco Bertone$^{1,3}$ and Joseph Silk$^{1,3}$\\
$^{1}$Denys Wilkinson Building, 1 Keble Road, Oxford OX1 3RH,
United
Kingdom\\
$^{2}$Instituto Superior T\'ecnico, Centro Multidisciplinar de
Astrof\'\i sica, Av. Rovisco Pais,
                1049-001 Lisboa, Portugal\\
$^{3}$Institut d'Astrophysique de Paris, F-75014 Paris, France}
\date{Draft version \today}
\pagerange{\pageref{firstpage}--\pageref{lastpage}}
\pubyear{2001}

\begin{document}

\maketitle

\label{firstpage}

\begin{abstract}
If the Milky Way is populated by Weakly Interacting Massive
Particles (WIMPs) as predicted by cosmological models of the
large-scale structure of the universe and as motivated by
supersymmetric models of particle physics (SUSY),
the capture of high-mass WIMPs by the Sun would affect the
temperature, density and chemical composition  of the solar core.
This is because WIMPs  provide an alternative mechanism for
transporting the energy of the core, other than radiative
transfer. Helioseismology provides a means for an independent test
of the validity of the WIMP-accreting solar models.
We use the sound speed and the density profiles inferred from the
helioseismic instruments on the Solar and Heliospheric Observatory
(SOHO) to discuss the effect of WIMP accretion and annihilation on
the evolution of the Sun. The WIMP transport of energy inside the
Sun is not  negligible for WIMPs with a mass smaller than $60\;
\rm GeV$ and annihilating WIMPs with $\langle \sigma_a
v\rangle\sim 10^{-27} \rm cm^3/sec$. WIMP-accreting models with
WIMP masses smaller than $30\;\rm GeV$ are in conflict with the
most recent  seismic data.
We combine our new  constraints  with the analysis of predicted
neutrino fluxes from annihilating WIMPs in the solar core. Working
in the framework of the Minimal Supersymmetric Standard Model
and considering the neutralino as the best dark matter particle
candidate, we find that supersymmetric models, consistent with
solar seismic data and with recent measurements of dark matter
relic density, lead to a measured muon flux on Earth in the
range of 1 to $10^4 \rm km^{-2}\; yr^{-1}$, for neutralino masses
between 30 and 400 GeV.  The local change of the solar core
structure combined with the increasing accuracy of solar models
and the increased sensitivity of future neutrino
telescopes presents a clear and distinctive seismic signature that
will enable
 us to set strong independent constraints on the physical
properties of dark matter particles.
\end{abstract}



\section{Introduction}
The dark matter problem has been an important
unresolved problem in astrophysics for several decades. It has
motivated much activity in cosmology, in both theoretical
and observational areas, leading to significant progress  in our
understanding of the evolution of the universe.
 To agree with the measured abundances of helium,
deuterium and lithium, the baryonic content of the Universe
$\Omega_b$
must satisfy $\Omega_b=0.04\pm 0.01$.
The dark matter is
 dominated by non-baryonic relic particles created
in the early stages of the Big Bang,
 amounting to $\Omega_m= 0.3\pm 0.1$.

The best-motivated non-baryonic dark matter
candidates are
 Weakly Interacting Massive Particles (WIMPs), which are
stable particles,  neutral and weakly interacting with ordinary
matter.
WIMPs were copiously produced in the early universe
through their weak interactions with other forms of matter and
radiation. As the universe expanded and cooled, their number
density  became too low for the annihilation processes
to keep up with the Hubble expansion rate. A relic population of
WIMPs should exist. Lee \& Weinberg (1977) showed that if such a
stable particle exists, its relic abundance is $\Omega_x h^{2}
\simeq 3\times 10^{-27} cm^3 s^{-1}/ \langle\sigma_a v\rangle$,
where $\langle\sigma_a v\rangle$ is the thermally-averaged product
of annihilation cross-section and relative velocity. A particle
which interacts with baryonic matter within the range of the weak
interactions will lead to $\Omega_x$ of the order of unity,
 in agreement with cosmological  determinations.
Specifically,  a recent analysis (Melchiorri and Silk 2002)
 points to $\Omega_x h^2 = 0.12 \pm
0.04$, thereby allowing us
to restrict ourselves to study the case of annihilating WIMPs with
annihilation  cross-section, $\langle\sigma_a v\rangle\sim
10^{-27} cm^3 s^{-1}$.

In this paper we choose as our favorite dark matter candidate the
so-called neutralino, arising in supersymmetric (SUSY) extensions of the
Standard Model of electroweak interactions. More precisely, we
work in the framework of the Minimal Supersymmetric Standard Model
(MSSM) as implemented by Gondolo et al. (2001). The neutralino is
expected to be the Lightest Supersymmetric Particle (LSP) in the
MSSM (Ellis 2001), almost independently of the  further simplifying
assumptions one has to make to reduce the huge number of free
parameters (63) contained in the MSSM.

 Neutralinos or other WIMP
candidates have not as yet been detected  in accelerators,
although LEP measurements set a lower bound of about 50 GeV on the
WIMP mass (Baltz and Gondolo 2001).
If WIMPs populate the halo of the Milky Way, then they can be
detected either directly in  low-background laboratory detectors
or indirectly via observation of anomalous cosmic-ray antiprotons,
positrons, and gamma rays from WIMPs that have annihilated in the
Galactic halo (Silk \& Srednicki 1984).
Furthermore, WIMPs may be captured in the Sun (Press and Spergel 1985;
Silk, Olive and
Srednicki 1985) or in the Earth (Freese 1986;
Gaisser, Steigman and S. Tilav 1986;
Krauss, M. Srednicki and F. Wilczek 1986;
Gould, Frieman and Freese 1989)
and annihilate, thereby producing high-energy neutrinos.
Annihilations balance capture,
leading to an  an equilibrium concentration in the solar core
 ( Gilliland, Faulkner, Press  and  Spergel 1986;
Krauss, Freese , Spergel, Press  1985;
Jungman, Kamionkowski  and Griest 1996;
Lopes, Silk and Hansen 2002). We henceforth only consider solar captures
 since the resultant neutrinos dominate the terrestrial muon detector signal.

The accretion by the Sun of WIMPs from the galactic background
population leads to efficient mechanisms of transport of energy in
the solar interior: such WIMPs provide an additional process for
transferring energy of the deeper layers to the  external layers,
depending upon the scattering cross-section of the WIMPs on the
solar nuclei. By virtue of their long mean free paths, WIMPs
transport energy radially, and tend to produce an isothermal core.
WIMP-accreting solar models  that have a current relatively
low concentration number of WIMPs reduce the temperature at the
star's core and produce a reduction in the solar neutrinos
generated by some nuclear reactions of the pp chain and CNO cycle.

This leads to a reduction of solar neutrino counting rates
measured on Earth. If such WIMPs exist, a decrease of the solar
neutrino fluxes arising from the existence of a WIMP isothermal
core could be successfully measured in the coming years by future
solar neutrino experiments (Lopes \& Silk 2002). Furthermore, if
heat transport by WIMPs is significantly changing the structure of
the solar core, a particular signature of this peculiar Sun's core
should be expected in  solar seismic data (Lopes, Silk and Hansen
2002).

The seismic diagnostics  of the Sun's interior have been for many
years the most important constraint on  the internal thermodynamic
structure of the Sun. Indeed, such research  has led to
significant improvements in  the microphysics such as an update of the
equation of state and the opacity calculations, and to a better
determination of specific cross-sections of the $pp$ chain
(Turck-Chi\`eze \& Lopes 1993; Christensen-Dalsgaard {\it et al.}
1996; Turck-Chi\`eze, Nghiem, Couvidat \& Turcotte 2001; Bahcall,
Pinssonneault, Basu 2001; Provost, Berthomieu \& Morel 2000). This
research, motivated by helioseismology, has led to a significant
improvement of the so-called Standard Solar Model. In
addition, through helioseismic inversion diagnostics, it is
possible to build solar models that evolve in an halo of WIMPs,
which are consistent with oscillation data and hence predict the
annihilating neutrino flux. Furthermore, there exist
WIMP-accreting solar  evolution
 models with different masses, scattering cross-sections, and
annihilation cross-sections, for which the present solar structure
is inconsistent with solar seismic data. The present article
discusses the validity of such models and the independent
method of diagnostic that seismology can provide
for probing the SUSY parameter space.

\section{The evolution model of the Sun}
Application of Newton's laws to our Galaxy tell us that the
luminous disk and bulge must be immersed in a dark halo with a
local density of 0.3 GeV cm$^{-3}$ and that dark matter particles
move with velocities comparable to the local circular speed.
Additional theoretical arguments suggest that the velocity
distribution of these particles is locally nearly isotropic and
nearly a Maxwell-Boltzmann distribution. When a WIMP enters a
star, it may interact with nuclei and lose enough kinetic energy
to be trapped by the gravitational potential well. The evolution
of the Sun is performed  within a sea of WIMPs. The WIMP gas tends
towards thermalization with baryonic  matter with a time-scale
much shorter than the time scale of stellar evolution. The WIMP
spatial distribution is then simply the barometric equilibrium
density, i.e., near a Gaussian with a typical length scale of
$r_x\sim 0.13\sqrt{\left(1GeV/m_x\right)}\; R_\odot$. The more
massive are the WIMPs, the more  they are  concentrated in a very
small region within the core of the star. The trapped WIMPs supply
another means of transferring energy from the energy producing
core to the outside, and thus supplement the usual transfer by
photon diffusion (Lopes, Silk and Hansen 2002). WIMPs might have
spin-independent (scalar) interactions in which case they would
interact with all chemical elements in the Sun, or they might have
only  spin-dependent (axial) interactions in which case they would
interact essentially only with hydrogen.

 In fact, at present, direct laboratory searches are   restricted to
SUSY particles with scalar interactions. For this reason, the
research reported  in this work will focus only on this type of
interactions. Current detectors (DAMA, CDMS, UKDMC) are sensitive
to scattering cross-sections $\sigma_s \gapprox 10^{-42}\rm cm^2$,
with one to two orders of magnitude improvement possible in the
near future. Future detectors (e.g. GENIUS, Cryoarray) plan to be
sensitive down to $\sigma_s \gapprox 10^{-45}\rm cm^2$. The
WIMP-accreting solar models are computed in a way similar to the
standard solar model, the only difference being the existence of
an alternative mechanism of transport of energy supplemented by
the presence of WIMPs.

As usual, in stellar evolution through the main sequence, we assume
that the star is in hydrostatic equilibrium and is spherically
symmetric, and that effects of rotation and magnetic fields are
negligible. The evolution of the star starts on the pre-main
sequence, 0.05 Gyr from the ZAMS. The solar structure and
evolution are calculated starting from an initially homogenous
star with a given composition. A Henyey method is used to solve
the system of nonlinear differential equations describing the
stellar structure (Morel 1997). Starting with a standard
primordial chemical composition the present solar luminosity and
radius is reached at its present age 4.6 Gyr, by readjusting the
initial helium abundance and the mixing length parameter (Lopes,
Silk and Hansen 2002).
It  follows that the final WIMP-accreting solar model is very
similar to the solar standard model, the difference between the
two models possibly  being identifiable via seismic diagnostics.
The transport of energy by WIMPs is strongly dependent on the mass
and scattering cross-sections.

In summary, the heat transport is optimized for
$\sigma_{s}\sim\sigma_{c}$ when the WIMP scale height is roughly
equal to its mean free path. $\sigma_c$ is a natural geometrical
scattering cross-section, depending on the proton mass $m_p$ and
the radius and mass of the star, $\sigma_c=m_p/M \;R^2=8\times
10^{-36}\;\rm cm^2$. In order to be effective in heat transport,
the WIMPs must have mean scattering cross-section per baryon in
the range of $10^{-43} \rm cm^2 \lapprox \sigma_{s} \lapprox
10^{-33} cm^2$, depending upon the annihilation cross-section and
mass of the WIMP. The transport of energy by WIMPs falls rapidly
outside this range, and it becomes very difficult to test this
effect on the solar structure against the solar seismic data. At
higher cross-sections the energy is transported locally and the
conductivity falls as $\sigma_c/\sigma_{s}$. At lower cross-sections
the conductivity falls as $\sigma_{s}/\sigma_c$ and in
addition only a fraction are captured by the Sun (Lopes, Silk and
Hansen 2002).

\section{Seismic diagnostic of the solar interior}
The WIMPs are thermalized within the solar core and are on
Keplerian  orbits around the solar center, interacting through
elastic scattering with solar nuclei, such as hydrogen and helium,
thereby providing an alternative mechanism of energy transport
other than radiation. The result is a nearly flat temperature
distribution, leading to an isothermal core. Consequently, the
central temperature is reduced. This reduction of temperature has
two main consequences: since central pressure support must be
maintained, the central density is increased in the WIMP-accreting
models, and since less hydrogen is burnt at the centre of the Sun,
the central helium abundance and the central molecular weight are
smaller than in standard solar models. The increase of the central
density and hydrogen partially offset the effect of lowering the
central temperature in the central production of energy. In fact,
this is the reason why minor changes are required to the initial
helium abundance and the mixing-length parameter in order to
produce a solar model of the Sun with the observed luminosity and
solar radius (Lopes, Silk and Hansen 2002). This readily leads to
a balance between the temperature, $T$, and the molecular weight,
$\mu$, in the core, leading to the peculiar profile of the  square
of the sound speed, $c_s^2\propto T/\mu$, and the density,
$\rho\propto \mu/T$, within the solar interior. This seems to be
the case for most of the WIMP-accreting solar models.

Figure 1 illustrates the differences between the radial profile of
the square of the sound speed and the radial profile of density
for WIMP-accreting  solar models and the Sun, as predicted by the
solar standard model theory. In the same figure, we illustrate the
square of the sound speed and the density as inferred for the
present Sun by using the data of Global Oscillations at Low
Frequency (GOLF) and Michelson Doppler Imager (MDI) experiments.
The average mean hydrostatic structure of the Sun has been
obtained by an optimally localized averaging inversion method  by
Kosovichev (1999). The method for independently determining  the
radial dependencies of the sound speed and density also yields the
radial dependence of the first adiabatic index or the chemical
composition. All the different methods of inversion are extremely
sensitive to the quality of the frequency measurements, and  very
accurate seismic data for the low-degree acoustic modes is
necessary for a precise inversion. The nuclear region is probed by
as many as 120 acoustic modes that are significantly influenced by
the turbulence, non-adiabatic effects and magnetic field
perturbations at the surface layers (Lopes \& Gough 2001).
Nevertheless, the long duration of continuous measurements has
reduced the uncertainty related with dynamics of the outer layers.

 In particular the global modes,
dipole modes and quadruple modes measured by the  GOLF experiment
(Bertello {\it et al.} 2000), coupled with high-degree modes
obtained by MDI  experiment data (Rhodes {\it et al.}
1997), enable us to significantly  constrain the central region of
the Sun, where the presence of WIMPs could be detected. However,
the inversion of the sound speed seems to be in better agreement
with the SSM than the inversion of density. This is due to the
fact that this inversion uses only acoustic modes. Naturally, in
such a case it is adequate to infer the sound speed profile rather
than the density profile in the nuclear region. To successfully
obtain the same level of accuracy for the density profile as in
the case of the square of the sound speed profile, it is necessary
to use gravity modes, which have not yet  been unequivocally
detected  (Turk-Chi\`eze {\it et al.} 2002).

Obviously, another reason for the density difference could come
from the physics of the solar standard model, and not only from
the density profile inversion. However, this is more unlikely
because the sound speed difference is very small, reenforcing the
view that the physics in the solar nuclear region is already
described with the necessary accuracy.

Even if the answers to some questions about the inversion are still
unclear, it seems very likely that there exists a class of
annihilating WIMPs with annihilation cross-section $\langle
\sigma_a v \rangle$ of the order of $10^{-27}\; \rm cm^{3}/sec$
and relatively small masses that are excluded by the present
seismic results. WIMPs with masses smaller than 60 GeV and
scattering cross-sections between $10^{-38} \rm cm^2$ and
$10^{-40}\rm cm^2$ seem to significantly modify  the profile of
the sound speed near the core.

This result is also reenforced by the density profile inversion.
If we accept   these  results, both inverted quantities reject the
existence of  WIMPs with masses smaller than $30 \; \rm GeV$ in
the proposed scattering cross-section range. It follows that the
acoustic spectrum of the previous WIMP-accreting solar models is
incompatible with the observed solar spectrum measured by the SOHO
seismic experiments. Usually, the solar model of the present Sun
which best reproduces the observed acoustic spectrum is referred
as the solar seismic model. In particular the solar standard model
leads to the best representation of the solar seismic model but it
is not the only solution.

\section{The neutrino flux of annihilating WIMPs}
Model-independent predictions can be made for neutrinos from the
centre of the Sun, where neutralinos may have been gravitationally
trapped and therefore their density enhanced. Today, the rate of
change of the number of neutralinos between capture and
annihilation in the Sun  is in equilibrium. As they annihilate,
many of the possible final states produce, after decays and
hadronization, energetic neutrinos which propagate out from the
interior of the Sun. In particular, the muon neutrinos are useful
for indirect detection of neutralino annihilation processes, since
muons have quite a long range in a suitable detector medium like
ice or water. They can be detected through their Cherenkov
radiation after having been produced at or near the detector.
Detection of neutralino annihilation into neutrinos is one of the
most promising indirect detection methods, and it will be subject
to extensive experimental investigations in view of the new
neutrino telescopes such as AMANDA, ICeCube, Baikal, BAKSAN,
MACRO, NESTOR and ANTARES planned or under construction (Halzen
1997). The neutrino-induced muon flux may be detected in a
neutrino telescope by measuring the muons that come from the
direction of the centre of the Sun or Earth. The energy of these
muons will typically be between 1/2 and 1/3 of the neutralino
mass, so they will be much more energetic than ordinary solar
neutrinos. These neutrinos have energies of the order of a GeV,
well above the energy of solar neutrinos which is of the order of
MeV.

To investigate this question, we concentrate our attention on
supersymmetric dark matter and
evaluate the expected flux of neutrinos (and consequent muon
fluxes in detectors) from annihilating neutralinos in the solar
core. The numerical results discussed in this work are obtained
in the framework of the Minimal Supersymmetric Standard Model
as implemented in
the DarkSUSY code (DMSSM; Gondolo et al. 2000), which takes
into account the  most recent particle physics constraints, such
as the LEP lower bounds on the lightest Higgs and chargino
masses. We extended our analysis to some benchmark points of a
different supersymmetric scenario, the  Constrained Minimal
Supersummetric Standard Model (CSSM; Ellis et al. 2000).
In this work we are mainly concerned with WIMP
candidates, such as the
neutralinos capable of producing changes in the structure of the
solar core which can be tested against the solar seismic model.
Indeed, even after all the particle accelerator and relic abundance
constraints are taken into account, there are large numbers of
SUSY models which can produce neutralino dark matter.
It is in this large parameter space of candidates proposed
by the different extensions of the standard model of particle
physics that the Sun can provide another independent
diagnostic of the SUSY parameter space.

One should be aware that the choice of nuclear form factors needed
to compute neutralino-nucleus elastic scattering, as well as other
specific quantities related to hadronic physics which relate
quarks/gluons with nucleons and also the quantities related with
the step from nucleons to nuclei, are at best approximate. This
is the reason why detailed processes such as scattering cross-sections
are difficult to obtain with any generality. A more
sophisticated treatment would, however, change the values by much
less than the spread due to the unknown super-symmetric
parameters. Indeed, our understanding of SUSY models is still
developing, so predictions of annihilation rates in the early
Universe, and thus relic neutralino densities may
require modification. More significantly, another source of
incertitude in modeling the incoming flux of annihilating neutrinos
is related with the poor description of the solar core usually
assumed in the computations. However, the present and future
capability of solar seismology will provide the means to
reduce this source of error.

Future solar seismic experiments will be able to detect deviations
of order $10^{-5}$ from the luminosity predicted in the solar
standard model.  If the microphysics of the solar standard model
is understood with the necessary precision, then at this level of
accuracy a large portion of the supersymmetric parameter space would
be ruled out by means of the seismic diagnostics. We show in Fig.~2 how this
analysis would affect the expected muon flux, induced by high energy
neutrinos from neutralino annihilation in the solar core, in the two
different supersymmetic scenarios discussed above: DMSSM and CMSSM.
A wide portion of the models obtained in DMSSM, namely the models
leading to the highest fluxes, would be probed at  $10^{-5}$ accuracy
using solar seismic data. The benchmark points for CMSSM  lie
outside most of the parameter space where solar seismic
data are sensitive enough
to the neutralino annihilation luminosity. The SUSY
parameter space that is constrained by helioseismology will be
discussed in more detail in a future paper (Bertone, Lopes, Sigl
\& Silk 2002, in preparation).

\section{Discussion and Conclusions}
 Although
important questions  still remain that need to be addressed
regarding the structure of the Sun, such
as the asymmetric macroscopic motions in the core, the dynamical
effects in the nuclear reaction rates and the chemical abundances
in the nuclear region (Turck-Ch\'eze {et al.} 2001), seismic
analysis has been a powerful  diagnostic tool for the Sun's
interior. This has led us to constrain the WIMP parameters based
on the proposition that the present standard solar model is an
accurate approximation of the observed Sun.

The WIMP-accreting solar models of this article are not certainly
an unequivocal constraint on the SUSY parameters. If we had
modified  some of the solar parameters within the error bars, such
as  for example the age of the Sun, we would probably have found
other solutions. However, the general results, as far as the WIMPs
are concerned, would be qualitatively the same.  The results
discussed in the article highlight the powerful tool that solar
seismology represents in order to constrain the WIMP parameters
within the expected experimental values. Such results strongly
favor the non-existence of annihilating WIMPs with $\langle
\sigma_a v \rangle\sim 10^{-27}\rm cm^3/sec$, masses lower than 60
GeV, and with scalar cross-sections from $10^{-40}\;\rm cm^2$ to
$10^{-38}\;\rm cm^2$.

It
should be noticed that the resolution in the inner core is
relatively poor; about $0.05R_\odot$, which is insufficient to
detect small discontinuities, probably due to the WIMP isothermal
core. This certainly justifies further searching for gravity
modes. The modelling of WIMP-accreting solar models is central to
the prediction of the muon flux that is expected to be measured by
the neutrino telescopes such as AMANDA, ICeCube, Baikal, BAKSAN,
MACRO, NESTOR and ANTARES (Halzen 1997 and references therein).
Assuming that the WIMP is a neutralino and the density of
neutralinos is in agreement with CMB and large-scale structure
observations, the prediction of the muon flux is within the range
of $1$ to $10^{4}$ $\rm km^{-2}yr^{-1}$, for neutralinos with
masses between 60 GeV and 400 GeV.  Theoretically, it is the low-degree
gravity modes that are the most sensitive to the conditions of the
nuclear region, at present the only region where substantial
deviations from the standard models occur. Indeed, accreting-WIMP
solar models have gravity modes period spacing that are markedly
different from that of other solar models, at least for the case
of smaller WIMP masses. The asymptotic period spacing of
high-order gravity modes of low degree is proportional to an
integral of the ratio of the buoyancy frequency by the radius, and
its present value for the solar standard model is of the order of
$36\; min$ (Lopes \& Turk-Chi\`eze 2002, in preparation).
Following the period separation between modes of the same degree
and consecutive radial-order, the period separation is expected to
be relatively smaller than in the case of solar standard models,
by as much as a few percent for the WIMP-accreting solar models
discussed in this work. It is the sensitivity to the buoyancy
that can be used to invert the density in the very deep layers of
the solar core. Therefore, gravity mode observations hold the
promise of a sensitivity test, although their current
interpretation is difficult between the possible candidates
detected (Turck-Chi\`eze {\it et al.} 2002). Furthermore, the
gravity mode data reduction is done assuming that the Sun is well
represented by the solar standard model, which could be a major
difficulty in terms of detection, if dark matter changes the
structure of the solar core.

In any case, the physical processes
presented here are meant to be indicative of what one might expect
for realistic WIMP-accreting models. Furthermore, since the
physics of solar models, the seismic data reduction, as well as
the SUSY model predictions are themselves evolving, the detailed
model results quoted here should be taken as indicative of the
general order of magnitude of  expectations for muon fluxes.

\section*{Acknowledgments}

The authors wish to thank P. Morel for using the CESAM code, and
P. Gondolo, J. Edsjo, L. Bergstrom, P. Ullio, E. A. Baltz,
 for using the darkSUSY code. Thanks also to S. Hansen, G. Sigl
and S. Turck-Chi\`eze for stimulating discussions on the
physics of super-symmetric models of particle physics and solar
models. IPL is grateful for support by a grant from Funda\c c\~ao
para a Ci\^encia e Tecnologia.

\label{lastpage}

\newpage
\onecolumn

\begin{figure*}
\centerline{
\psfig{file=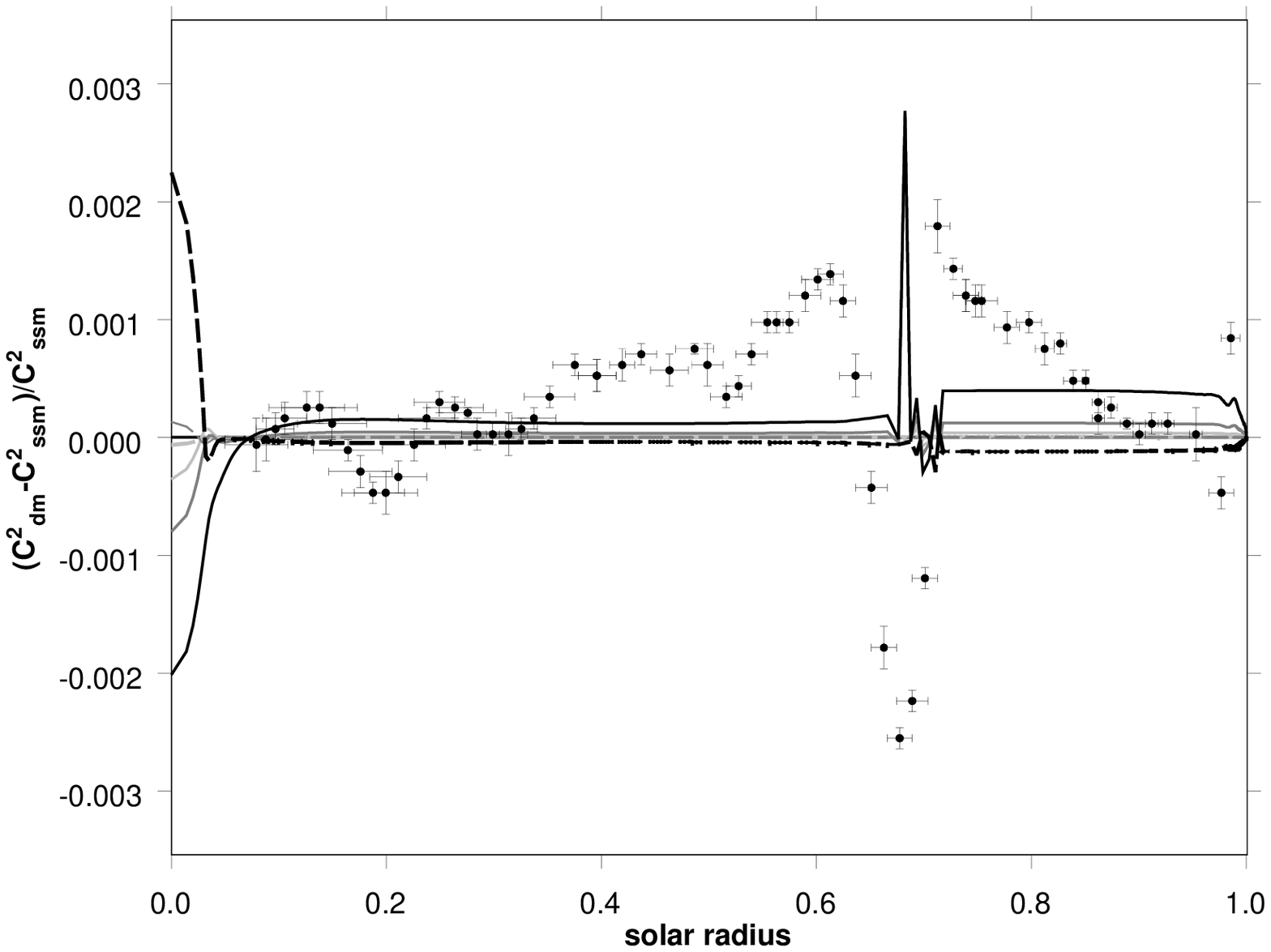,width=14cm,height=10.0cm}}
\vspace{-1.5cm} \centerline{
\psfig{file=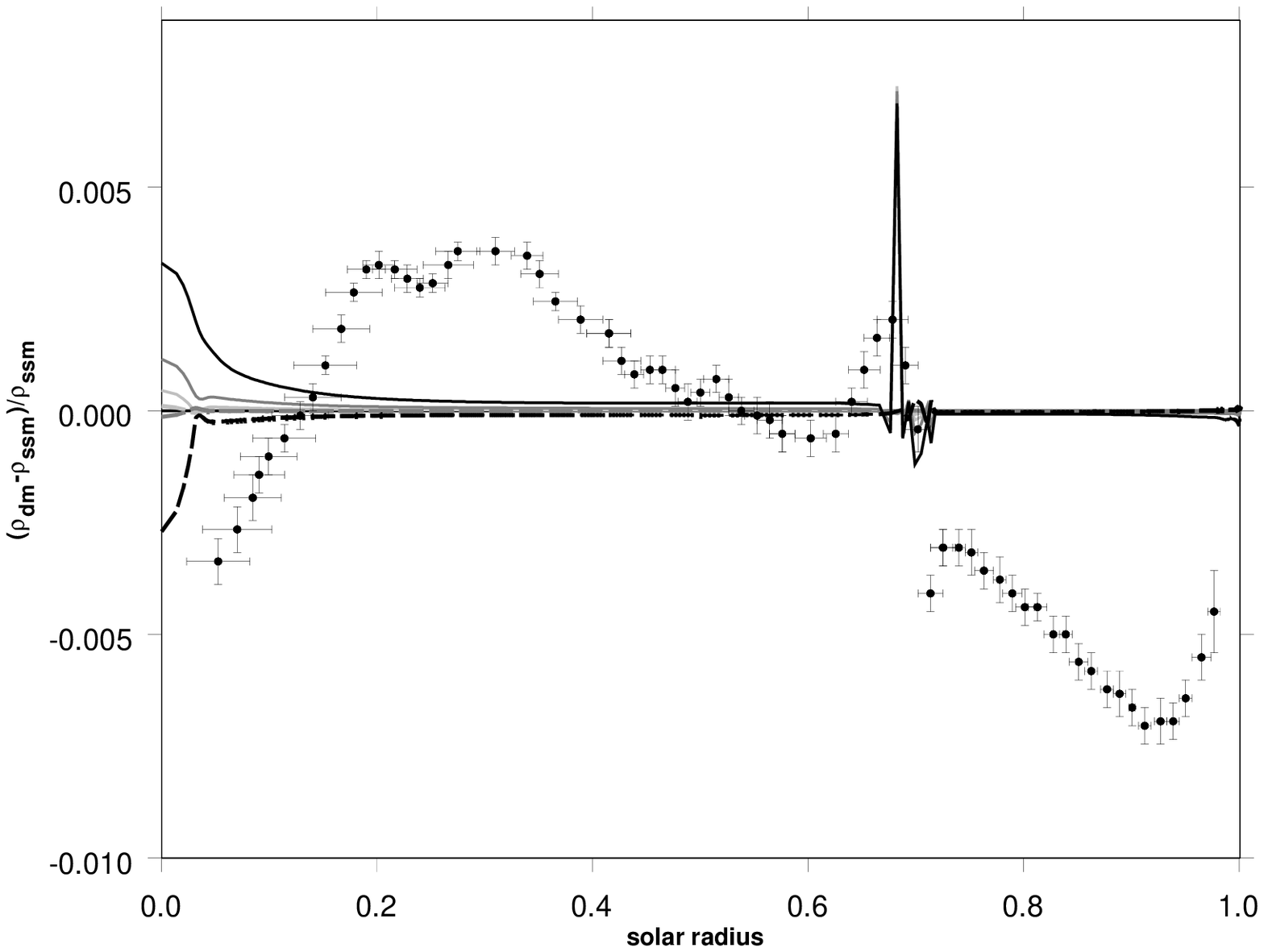,width=14cm,height=10.0cm}}
\vspace{-0.5cm} {\bf Figure 1:} The relative differences between
the square of the sound speed ({\bf a}) and the density ({\bf b})
of the standard solar model and solar models, evolved within an
halo of WIMPs. The continuous curves correspond to models with
WIMP masses of 15 GeV, 30 GeV and 60 GeV and with
annihilation rate of $10^{-27}cm^3/s$,
and scalar scattering
cross-section of $10^{-38} cm^2$ or $10^{-40} cm^2$. The curves
are as follows: $m_x\sim 60\; GeV$ $\sigma_{s}\sim 10^{-38} cm^2$
(black curve), $m_x\sim 60\; GeV$ $\sigma_{s}\sim 10^{-40} cm^2$
(dark grey curve), $m_x\sim 30\; GeV$ $\sigma_{s}\sim 10^{-38} cm^2$
(light grey curve),$m_x\sim 30 \; GeV$ $\sigma_{s}\sim 10^{-40} cm^2$
(dashed light grey curve), $m_x\sim 15\; GeV$ $\sigma_{s}\sim 10^{-38}
cm^2$ (dashed dark grey curve) and $m_x\sim 15\; GeV$ $\sigma_{s}\sim
10^{-40} cm^2$ (dashed black curve). The curve with error bars represents
the relative differences between the square of the sound speed in
the Sun (as inverted from solar seismic data) and in a standard
solar model (Kosovichev {\it et al.} 1997; 1999). The horizontal
bars show the spatial resolution, and the vertical bars are error
estimates.
 \label{fig:1}
\end{figure*}

\clearpage

 \begin{figure*}
\centerline{\psfig{file=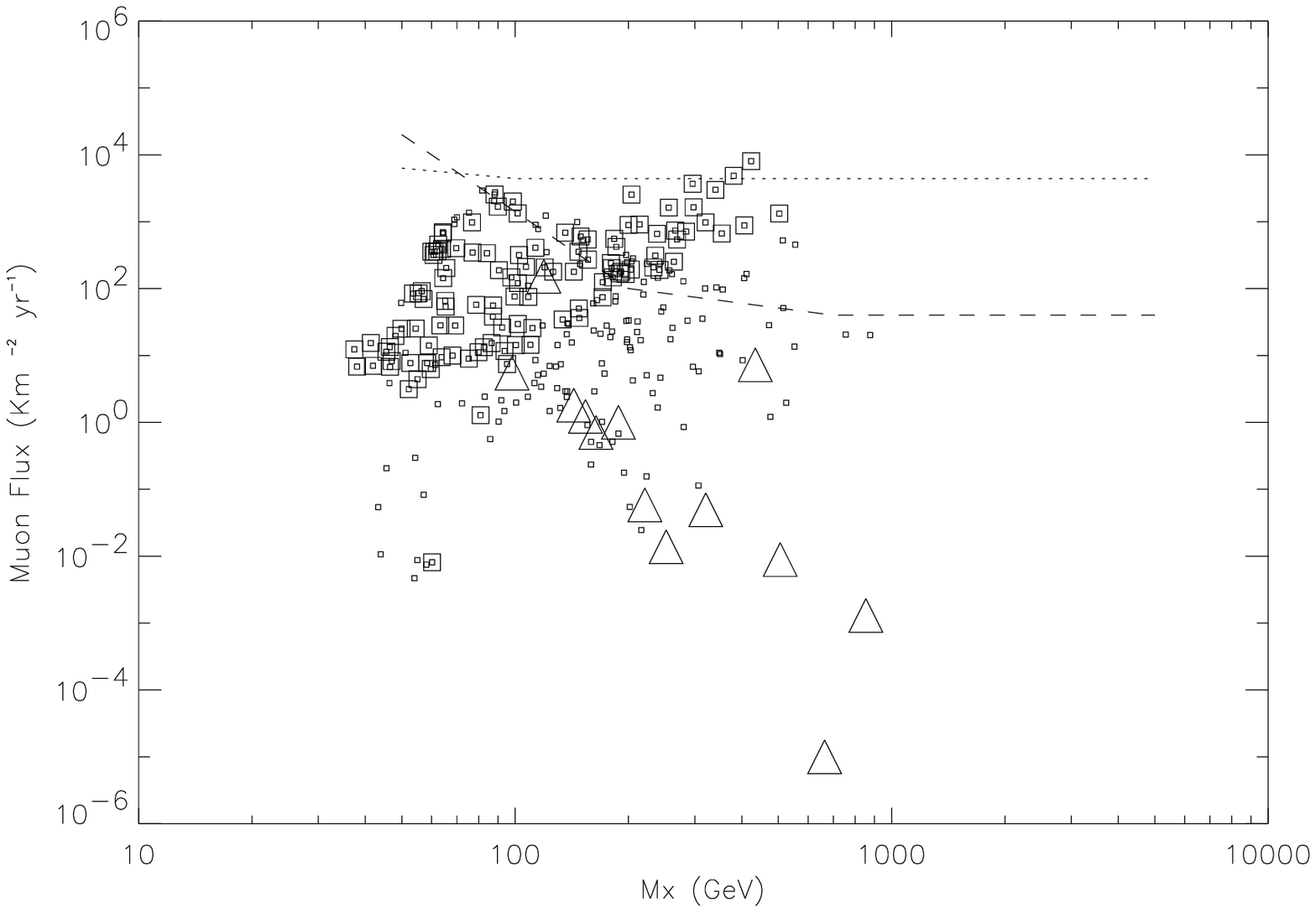,width=14cm,height=10.0cm}}
 {\bf
Figure 2:}
Predicted neutrino-induced muon flux produced by neutralino annihilation
in the Sun. Small squares correspond to models obtained with the
DarKSUSY code (DMSSM; Gondolo et al. 2000), triangles correspond
to selected benchmark points of the Constrained Minimal
Supersymmetric Standard Model (CMSSM; Ellis et al. 2000).
Big squares are used to highlight models leading to a local variation
of luminosity of the solar core larger than $10^{-5}$ (which could
thus be potentially probed by upcoming solar seismic observations).
The dotted and dashed curves represent the current limit sensitivity
of MACRO and Icecube experiments.
 \label{fig:2}
\end{figure*}

\end{document}